\documentclass[prl,noshowpacs,english,superscriptaddress,twocolumn]{revtex4-1}
\usepackage[colorlinks,linkcolor=blue,anchorcolor=blue,citecolor=blue,filecolor=blue,menucolor=blue,runcolor=blue,urlcolor=blue,frenchlinks=blue]{hyperref}
\usepackage{amsmath,amssymb,amsthm,amsfonts,mathrsfs}
\usepackage{graphicx}
\usepackage{pdfpages}
\usepackage{appendix}
\usepackage{lipsum}
\usepackage{slashed}
\usepackage{dcolumn}
\usepackage{mathrsfs}
\usepackage{bm}
\usepackage{color}
\usepackage{lineno,hyperref}
\makeatletter

\newcommand{\Rmnum}[1]{\expandafter\@slowromancap\romannumeral #1@}
\AtBeginDocument{\let\LS@rot\@undefined}
\makeatother
\usepackage{tikz,pgf}
\usetikzlibrary{shadings}
\usepackage{pgfplots,pgfplotstable}

\begin{document}
	
	\title{A spectrum-based shortcut method for topological systems}
	
\author{Jian Xu}
\email{xujian$_$328@163.com}
\affiliation{College of Electronics and Information Engineering, Guangdong Ocean University, and Guangdong Engineering Technology Research Center of Intelligent Ocean Sensor Network and Equipment,Zhanjiang,guangdong, 524088, China}
\affiliation{Guangdong Provincial Smart Ocean Sensing Networks and Equipment Engineering Technology Research Center, Zhanjiang 524088, China}

	\author{Feng Mei}
	\affiliation{State Key Laboratory of Quantum Optics and Quantum Optics Devices, Institute of Laser Spectroscopy, Shanxi University, Taiyuan, Shanxi, 030006, China}

\author{Yan-Qing Zhu}
\email{zhuyq1992@gmail.com}
	\affiliation{Quantum Science Center of Guangdong-Hong Kong-Macau Great Bay Area, 3 Binlang Road, Shenzhen, China}
	\affiliation{Guangdong-Hong Kong Joint Laboratory of Quantum Matter, Department of Physics, \\and HK Institute of Quantum Science $\&$ Technology,\\ The University of Hong Kong, Pokfulam Road, Hong Kong, China}

	\date{\today}
	
	\begin{abstract}

The need for fast and robust quantum state transfer is an essential element  in scalable quantum information processing, leading to widespread interest in shortcuts to adiabaticity for speeding up adiabatic quantum protocols. However, shortcuts to adiabaticity for systems with more than a few levels is occasionally challenging to compute in theory and frequently difficult to implement in experiments. In this work, we develop a protocol for constructing shortcuts to adiabaticity through the multi-state Landau-Zener approach and a stricter adiabatic condition. Importantly, our protocol only requires a few pieces of information about the energy spectrum and just adjusts the evolutionary rate of the system. It means that our protocol has broad applicability to theoretical models and does not require increasing the difficulty of the experiment. As examples, we apply our protocol to state transfer in the two-level Landau-Zener model, the non-Hermitian Su-Schrieffer-Heeger (SSH) model and the topological Thouless pump model and find that it can speed up the manipulation speed while remaining robust to Hamiltonian errors.  Furthermore, based on the experimental friendliness of our findings, it can potentially be extended to many-body systems, dissipation cases, or Floquet processes. Overall, the proposed shortcut protocol offers a promising avenue for enhancing the efficiency and reliability of quantum state transfer protocols.

	\end{abstract}
	
	\maketitle
	
	\emph{{\color{blue}Introduction.}---} Adiabatic control over quantum states is of fundamental importance for quantum science. Generally, adiabatic processes in quantum systems are typically slow and steady, but this can lead to errors, perturbations, and decoherence accumulations that can even cause the system to escape from its confinement. To overcome this challenge, significant  efforts have been made over the past two decades to develop optimal protocol that speeds up the evolution process while maintaining the desired system within the adiabatic regime. Most of these shortcuts rely on the addition of auxiliary time-dependent couplings or specific time dependence of controlling parameters of a system\cite{Bergmann,Petr,Guery}, e.g., counterdiabatic driving\cite{Berry}, dynamical invariants\cite{Chen}, quasiadiabatic dynamics\cite{Martinez} and so forth \cite{YXDu2016}. 

Although shortcuts to adiabaticity are widely used in the context of few-level systems, they have not been well explored for multi-level systems, especially topological systems. The main obstacle is the difficulty in acquiring and employing the counterdiabatic term in multi-level systems, which requires  extensive knowledge of the full Hamiltonian spectruml properties and the ability to control highly non-local, infinite-range interactions. Recent progress has been made in systematically building approximate counterdiabatic terms using a variational method\cite{Sels}. This method has been applied to ground-state preparation\cite{Claeys,Passarelli,Prielinger,Barone}and state transfer\cite{Cepaite} in specific quantum spin systems. However, an exact and generic shortcut protocol for precise state manipulations in these systems has yet to be developed.

Recently, quantum state transfer via topological edge states has been proposed and studied in the topological phases of matter, as the edge modes are quite robust against  perturbation and disorder due to their topological nature in the bulk\cite{FMei2018,Qi,Yue2023}. Quantum state transfer via the topological edge states has been experimentally demonstrated with Rydberg atoms\cite{S. de}, semiconductor quantum dots\cite{Kiczynski}, and coupled waveguides\cite{Gabriel}. To date, speeding up this type of state transfer has been achieved through the special modulation function\cite{Brouzos,Palaiodimopoulos}, introducing additional coupling fields\cite{Felippo,Yunlan,Lijun} or by topological domain walls\cite{Zurita} in Hermitian cases. However, a proposal for enhancing state transfer via topological edge states, especially in non-Hermitian cases\cite{Cui}, remains an open question.

Similarly, Thouless pumping via adiabatic and periodic modulation, which is topologically protected, has been observed in various  artificial systems, including metamaterials\cite{Ke,Cerjan,Jurgensen1,You,Sun,Jurgensen2}, nitrogen-vacancy centre in diamond\cite{Ma,Boyers}, photonic systems\cite{Jurgensen} and ultracold atoms\cite{Lohse,Nakajima1,Lu} considering disorders \cite{Nakajima2}, interactions\cite{Walter}, or gains/losses\cite{Fedorova,Dreon}. Since its topological nature is robust only under adiabatic driving paths, optimizing the adiabatic process in practical implementations is crucial. Some shortcut schemes have been proposed, such as next to nearest neighbour hopping and the addition of a harmonic potential\cite{Schouten}, a family of finite-frequency protocols\cite{Malikis} and Bloch oscillations\cite{Liu}. But none are experimentally friendly. Therefore, proposing a feasible control protocol to optimize the speed and accuracy of the adiabatic passage remains a challenge.
	
In this work, we put forward a computationally efficient and experimentally friendly method for designing fast adiabatic protocols via a state transfer, which is in generic applicable to topological systems. The theoretical derivation for the adiabatic condition is first presented. We show that the adiabatic passage of parameter modulation can be determined through a simple calculation once the part of the eigenvalues of the system are obtained. More importantly, this method, which excludes auxiliary controls, will not increase experimental complexity. To demonstrate the advantages of our shortcut method, we apply this theory to investigate the two-level Landau-Zener model, the state transfer in the non-Hermitian SSH model and the topological Thouless pump models. Compared to common linear parameter modulation, our shortcut method provides optimal adiabatic passages.

	\emph{{\color{blue} Shortut method.}---} In general, a non-degenerate $N$-level non-Hermitian system under a time-modulation parameter $R(t)$  can be described simply as,
\begin{equation}
H_{N}(t)=\frac{1}{2}\left[\Omega+\delta(t)\right],
\end{equation}
with the time-independent non-diagonal component $\Omega_{ij}$ denoting the coupling between levels $E_i$ and $E_j$, where generically $\Omega_{ij}\neq \Omega_{ji}^*$ due to the non-Hermiticity. Meanwhile, we take the time-modulation term as a diagonal Hermitian matrix, i.e., $\delta=\text{diag}(\delta_1,\delta_2,...,\delta_N)$. Although many integrable models based on this have been explored, e.g. multi-state Landau-Zener models\cite{Chernyak2018,Chernyak2021,Suzuki} and multi-level Landau-Zener-Stückelberg-Majorana interference model\cite{Ashhab,Xiao,Ivakhnenko}, a generic method to analytically analyze the dynamics of this system is not yet known. The core of the difficulty is that an analytical theory must include full information about eigenenergies and eigenstates of the system, and the calculation of the eigenstates is usually difficult to obtain. However, in some cases, such as adiabatic transfer in this work, an analytical theory that includes only the calculation of eigenenergies has been deduced.

Since the Landau-Zener tunneling between any two levels is independent, without loss generality, we consider the adiabatic transfer in an effective two-level system with the Hamiltonian is given by,
\begin{equation}
\label{H_lz1}
H_{\text{eff}}(t)=\frac{1}{2}\left(
               \begin{array}{cc}
                 \delta(t) & \Omega_{0} \\
                 \Omega_{0}' & -\delta(t) \\
               \end{array}
             \right).
\end{equation}
Its instantaneous spectrum reads $E_{\pm}=\pm\sqrt{\delta(t)^2+\Omega_0\Omega_0'}/2$ with the corresponding eigenvectors,
\begin{equation}
\begin{split}\label{EV}
|\phi_+\rangle=\sin(\theta/2)|1\rangle+\cos(\theta/2)|0\rangle,\\
|\phi_-\rangle=\cos(\theta/2)|1\rangle-\sin(\theta/2)|0\rangle,\\
\end{split}
\end{equation}
where the mixing angle $\theta=\arccos \left(\delta/\sqrt{\delta^2+\Omega_0\Omega_0'}\right)$,  $|1\rangle=(1,0)^T$, and $|0\rangle=(0,1)^T$. Because the system is now non-Hermitian, we mention on its Hermitian conjugate, which satisfies
$ H_{\text{eff}}^{\dagger} |\tilde{\phi}_{\pm}\rangle=\tilde{E}_{\pm}|\tilde{\phi}_{\pm}\rangle$. The corresponding eigenvalue $\tilde{E}_{\pm}=E^*_{\pm}$ and $|\tilde{\phi}_{\pm}\rangle$ takes the same form as in Eq. \eqref{EV} by just replacing $\theta$ with $\tilde{\theta}$, where  $\tilde{\theta}=\arccos \left(\delta/\sqrt{\delta^2+\Omega^*_0(\Omega_0')^*}\right)$. Note that $\theta$ and $\tilde{\theta}$ are complex and the eigenvectors satisfy the bi-orthogonal relation, i.e., $\langle\tilde{\phi}_{\alpha}|\phi_{\beta}\rangle=\langle\phi_{\alpha}|\tilde{\phi}_{\beta}\rangle=\delta_{\alpha\beta}$.
	
For such a non-Hermitian system, the adiabatic condition is given by\cite{SM,SM2}
\begin{equation}\label{ABC}
\left|\frac{\langle \tilde{\phi}_{\mp}|\partial H_{\text{eff}}/\partial \delta|\phi_{\pm}\rangle}{(E_{\pm}-E_{\mp})^2}\frac{\partial \delta}{\partial t}\right| e^{\int_0^t \text{Im} (\Delta E)d\tau}\ll 1,
\end{equation}
with the band gap $\Delta E=E_{+}-E_{-}=\sqrt{\delta_0^2+\Omega_0\Omega_0'}$. The phase $e^{\int_0^t \text{Im} (\Delta E)d\tau}$ represents the cumulative effect of non-hermitian.  Eq.(\ref{ABC}) implies that the system evolution, either too rapid or too slow, will respectively be influenced by non-adiabatic effects and non-Hermitian effects.
Since $\partial H_{\text{eff}}/\partial \delta=\sigma_z/2$, we obtain a simpler relation,
\begin{equation}\label{C1}
|\langle \tilde{\phi}_{\mp}|\partial H_{\text{eff}}/\partial \delta|\phi_{\pm}\rangle|=\sin\frac{\theta+\tilde{\theta}}{2}.
\end{equation}
Considering that the non-Hermitian term remains invariant throughout the entire evolution process, $\sin\frac{\theta+\tilde{\theta}}{2}$ possesses a definite maximum value.  In addition, we could vary $\Delta E$ with respect to $t$, then we have 
\begin{equation}\label{C2}
\partial_t \Delta E=\delta\partial_t \delta/\Delta E.
\end{equation}
Substituting Eqs. \eqref{C1} and \eqref{C2} into Eq. \eqref{ABC}, and replacing $|\langle \tilde{\phi}_{\mp}|\partial H_{\text{eff}}/\partial \delta|\phi_{\pm}\rangle|$ with the maximum value of $\sin\frac{\theta+\tilde{\theta}}{2}$, while merging it into the adiabatic coefficient. We obtain a new adiabatic condition as
\begin{equation}
s=\left|\frac{\partial_t \Delta E}{\Delta E\sqrt{\Delta E^2- E_{\text{min}}^2}}\right|e^{\int_0^t \text{Im}(\Delta E)d\tau}\ll 1,
\end{equation}
where $ E_{\text{min}}=\sqrt{\Omega_0\Omega_0'}$ represents the minimum band gap, also known as an avoided crossings.
 Say $s$ is the coefficient to be determined, related to the total evolution time of the system. When $s\ll 1$, the system naturally satisfies the adiabatic condition as in Eq. \eqref{ABC}. In fact, realistic physical manipulations rely more on the parameter modulation, thus we prefer the following relation deforming from  $s$, i.e.,
 \begin{equation}\label{Rt}
 \partial_t \delta(t)=\left|\frac{s\Delta E\sqrt{\Delta E^2- E_{\text{min}}^2}}{\partial_t \Delta E}\right|e^{\int_0^t\text{-Im} (\Delta E)d\tau}.
 \end{equation} 
In this way, the control parameter $\delta(t)$ change slow near avoided corossings but fast away from them to make diabatic transitions weaker than the preconfigured adiabatic condition $s$ along the whole process.

In other words, we now provide a spectrum-based shortcut method (SBSM) that depends only on partial information of spectrum, which is friendly for topological systems, to speed up adiabatic passage. Therefore, we can now successfully address previous intractable systems that conventional schemes are powerless.
	 In the following text, we apply the above theory and show the numerical results in three different models, respectively, including the two-level Landau-Zener model, the 1D non-Hermitian SSH model and the Thouless charge pumping .
	
\emph{{\color{blue}Application I.}---} Firstly, consider a two-level model with a single avoided crossing.  In the bare basis, $|1\rangle=(1,0)^T$ and $|2\rangle=(0,1)^T$, Eq.(\ref{H_lz1}) becomes
\begin{equation}
\label{H_lz2}
H_{\text{eff}}(t)=\frac{1}{2}\left(
               \begin{array}{cc}
                 \delta(t) & \omega \\
                 \omega & -\delta(t) \\
               \end{array}
             \right).
\end{equation}
 Suppose $\delta(t)$ satisfies $\delta(0)=0$ and $\delta(-t)=-\delta(t)$. During the process $-t \rightarrow t$, if the evolution rate is sufficiently fast, the system will undergo non-adiabatic transitions. For example,  if $\delta(t)=v t$, Eq.(\ref{H_lz2}) becomes the standard Landau-Zener model.  To minimize non-adiabatic transitions during the evolution process, we now consider employing the USM proposed above.  Considering the spectrum $E_{\pm}^{lz}=\pm\sqrt{\delta(t)^2+\omega^2}/2$, and $\Delta E_{\text{min}}=\omega$, Eq.(\ref{Rt}) becomes $ \partial_t \delta(t)=s(\delta(t)^2+\omega^2)$. So based on USM,  we have  
  \begin{equation}\label{Rt_{lz}}
 \delta(t)=\omega \tan (s \omega t).
 \end{equation}

	\begin{figure}[http]\centering
		\includegraphics[width=8.4cm]{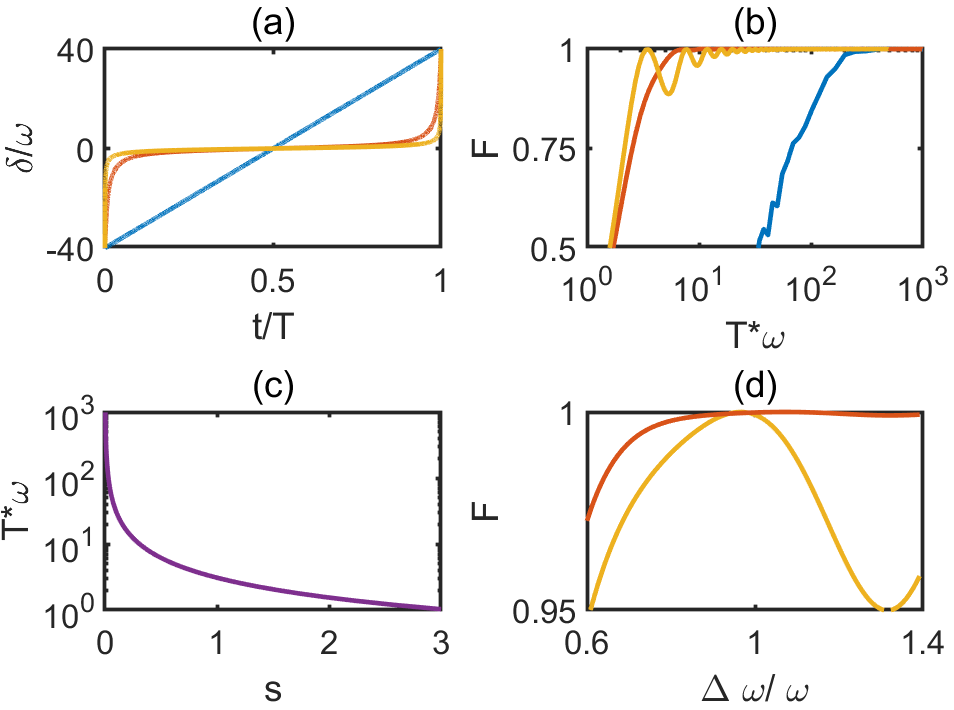}

		\caption{ Comparison of the linear modulation(blue line), FAQUAD (yellow line)  and SUSM(red line) (a) The curve of $\delta(t)$ and (b) the fidelity $F$ with different modulation methods. (c) The the total evolution time $T$ as a function of theadjacent adiabatic coefficient $s$ for SUSM. (d) The transfer fidelity vs $\omega$ variations for FAQUAD and SUSM in the case of $T*\Omega=18.6$.}
		\label{lz}
	\end{figure}
Afterwards,  we consider the system's excitation as it sweeps through avoided-crossing region. Suppose $\delta(T)=40\omega$ and the system starts from the $|\psi(-T)\rangle=|1\rangle$. when $t$  is varied from $-T \rightarrow T$,  there may be nonadiabatic transitions near the avoided-crossing. To characterize the non-adiabatic transitions, we utilize the fidelity $F=\langle \psi(T)|1\rangle$ to define the magnitude of non-adiabatic transitions. Fig. \ref{lz}(a) shows the curve of different methods and Fig.\ref{lz}(b) shows the fidelity $F$ as a function of evolution time $T$ under different methods. The adjacent adiabatic coefficient $s$ is shown in Fig.\ref{lz}(c).  Moreover, Fig.\ref{lz} (d) shows that  fidelity is a function of the coupling $\omega$ derived from the FAQUAD and our optimized method.  It can be observed that the SBSM significantly enhances the robustness with respect to the period $T$ and coupling strength $\omega$, while maintaining comparable speed to the fast quasiadiabatic modulation. For multi-level systems, robustness to parameter errors is the most critical evaluation criterion for control methods, as they must contend with the potential errors of many parameters during manipulation.

\begin{figure}[http]\centering
		\includegraphics[width=8.4cm]{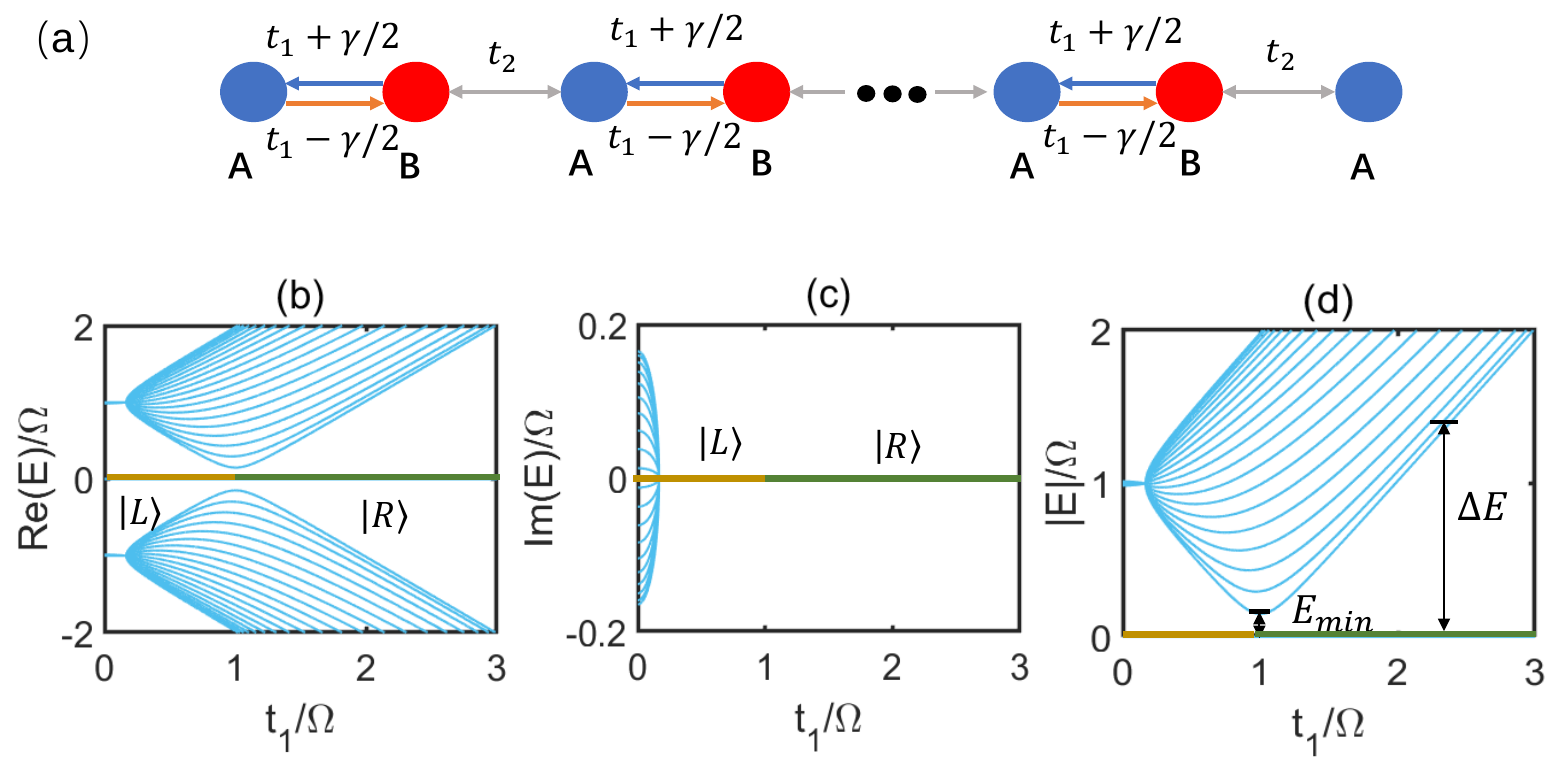}
		\caption{ (a) Schematic diagram of the 1D NH SSH model where a lattice chain consists of an odd number of lattice sites with each unit cell containing two subalttices labelling A and B, respectively. $t_1$($t_2$) denotes the intra(inter)-cell hopping while $+(-)\gamma/2$ is the NH intracell hopping from $B$($A$) to $A$($B$). (b),(c) The real and imaginary parts of energy spectrum. (d) The absolute value of the energy spectrum.  }
		\label{BDmode}
	\end{figure}

\emph{{\color{blue}Application II.}---} Next, we consider the 1D NH SSH model in the real space whose Hamitonian takes the form,
\begin{equation}
	\begin{split}
		H_1=\sum_{<i,j>}\left[(t_1+\gamma/2)c^{\dagger}_{i,A}c_{i,B}+(t_1-\gamma/2)c^{\dagger}_{i,B}c_{i,A}\right.\\
		\left.+t_2c^{\dagger}_{i,B}c_{j,A}+t_2c^{\dagger}_{j,A}c_{i,B}\right],
	\end{split}
\end{equation}
where a unit cell contains two sublattices labeled $A$ and $B$, $<i,j>$ denotes the nearest-neighbour unit cell and the hopping terms $H_\pm=t_1\pm \gamma/2 $, as depicted in Fig. \ref{BDmode}(a).

This model supports the non-Hermitian skin effect and its topological zero-modes are determined by the non-Bloch winding number as discussed in Ref. \cite{SYao2018}. To explore the state transfer via the edge state, we consider $H_1$ with $2N-1$($N$ is positive real number) lattice sites.  For this lattice chain, its topological zero-energy edge state is well-separated from the bulk states. Specifically, we find that there is one edge state $|L\rangle=|egg...g\rangle$ in the left end when $t_1\in [0,\sqrt{t_2^2+(\gamma/2)^2})$ and one edge state $|R\rangle=|gg...ge\rangle$ in the right end when $t_1\in (\sqrt{t_2^2+(\gamma/2)^2},3]$, as depicted in Fig. \ref{BDmode}(b-c). That is to say, as we tune $t_1$ from $0$ to $3$ adiabatically, there is a zero-mode edge state transferring from the left end to the right end, i.e. $|L\rangle\rightarrow|R\rangle$. Note that both the left and right edge states occupy the A-type sublattice in each unit cell, like in the Hermitian case\cite{FMei2018}.

\begin{figure}[http]\centering
	\includegraphics[width=8.4cm]{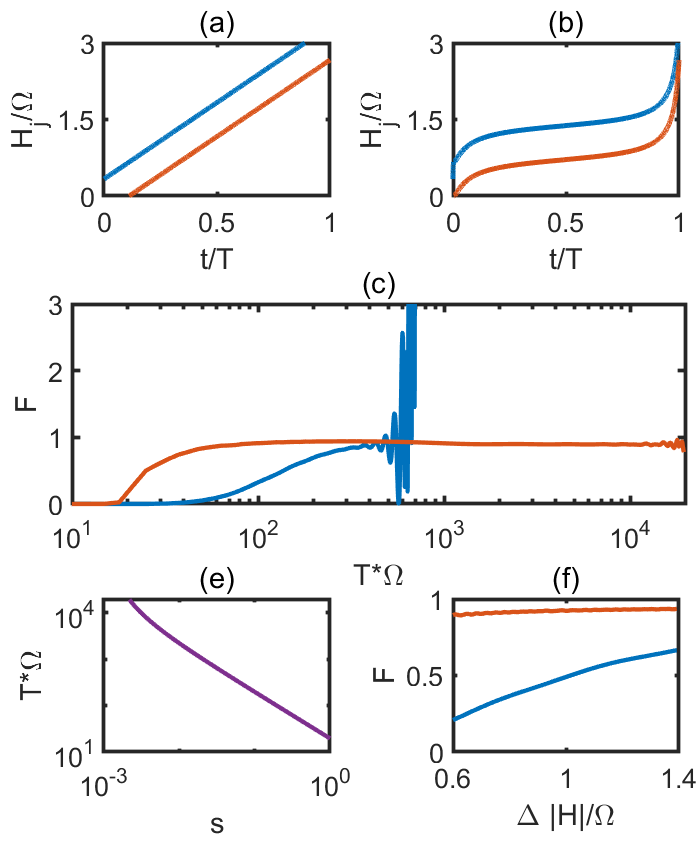}
	\caption{Comparison of the linear modulation and our optimized modulation, and the lattice size $N=20$. The $H_\pm$($H_+$: blue line, $H_-$: red line) vs $t_1/T$ for (a) the linear modulation and (b) the USE . The (c-d) show the transfer fidelity and the adjacent adiabatic coefficient $s$ as a function of $T*\Omega$ for the conventional linear modulation(blue line) and the optimal adiabatic modulation(red line) in the case of $\gamma=2\Omega/3$. (e) The transfer fidelity vs hopping term variations for the conventional linear modulation(blue line) and the optimal adiabatic modulation(red line) in the case of $T*\Omega=125$.}
	\label{fig2}
\end{figure}

In what follows, we mainly focus on the quantum state transfer via the topological edge states using our universal shortcut method.  When t is varied from $0\rightarrow3$, $|\psi(0)\rangle=|L\rangle\rightarrow|\psi(3)\rangle$. Because the upper branch and the lower branch of this system are symmetrical, only the absolute value of the energy spectrum needs to be determined, as shown in Fig. \ref{BDmode}(d). Fig. \ref{fig2}(a) shows the curve of $t_1(T)$ and also the linear modulation of $t_1$ while the optimal adiabatic modulation of $t_1(T)$ via our USM is depicted in Fig.\ref{fig2}(b). Note that the complex energy spectrum for this system [Fig. \ref{BDmode}(d)] can be measured in photonic/phononic crystals \cite{Bergholtz2021,WZhu2023} or through the non-Hermitian absorption spectroscopy method developed recently in cold atoms \cite{YXu2022,MCao2023}. Fig.\ref{fig2}(c) shows the fidelity $F=\langle R|\psi(3)\rangle$ as a function of evolution time $T$ under linear modulation and SUSM. The adjacent adiabatic coefficient $s$ is shown in Fig.\ref{fig2}(d). Moreover, Fig.\ref{fig2} (e) shows that  fidelity is a function of the hopping coefficient $t_1$ derived from the linear modulation and SUSM. Whether the system is Hermitian or non-Hermitian, our method can realize faster, high fidelity and more robust state transfer, compared with linear modulation. It is worth mentioning that our shortcut scheme can partly suppress the dissipation in the adiabatic process origined from the non-Hermiticity of the system.

\emph{{\color{blue}Application III.}---} By eliminating $\gamma$ term and adding an on-site term to $H_1$, we obtain the famous Rice-Mele model $H_{2}$, which is given by
\begin{equation}
	H_2=H_1(\gamma=0)+\sum_{i,m\in \{A,B\}} t_m c^{\dagger}_{i,m}c_{i,m},
\end{equation}
with $t_A=-t_B=\Delta(\phi)/2$. If we further set $t_1=t_0+\delta(\phi)$ and $t_2=t_0-\delta(\phi)$ with $(\delta,\Delta)=(\delta_0\cos \phi,\Delta_0 \sin \phi)$ [$\delta_0$ and $\Delta_0$ are periodic modulation parameters and $\phi\in [0,2\pi)$], we obtain the Thouless charge pump model. This topological pumping model reveals the first Chern number $C_1=\text{sgn}(t_0\delta_0\Delta_0)$ of the corresponding 2D system in the $(k,\phi)$ space, which has been well studied in theories and experiments \cite{DWZhang2018,Lohse, Nakajima1,Lu} . Considering the periodic condition in real space and after one pumping cycle, particles in the specific unit cell will be pumped to the next unit cell with a quantified change of polarization $\Delta P=P(\phi=2\pi)-P(\phi=0)=C_1$ where the lattice constant is unity. Here the polarization $P=\langle w(x)|\hat{x}-R|w(x)\rangle=\frac{1}{2\pi}\int _{-\pi}^{\pi} A_{k} dk $ with $|w(x)\rangle$ denoting the wannier state at unit cell $x=R$ in real space. The Berry connection being $A_k=\langle u_k(\phi)|i\partial_k|u_k(\phi)\rangle$ for the lower occupied Bloch band which leads to the associate Berry curvature $F_{k\phi}=\partial_{k}A_{\phi}-\partial_{\phi}A_k$. As we can see that $P$ is expressed though the integral of Berry connection in momentum space, leading to $\Delta P$ is proportional to the quantized $C_1$ by applying the Stokes' theorem during this pumping circle. i.e.,
\begin{equation}
	\begin{split}
	\Delta P&= \frac{1}{2\pi}\int _{-\pi}^{\pi}\left[A_{k}(\phi=2\pi)-A_{k}(\phi=0)\right] dk \\
	        &=\frac{1}{2\pi}\int _{-\pi}^{\pi}\left[A_{k}(\phi=2\pi)-A_{k}(\phi=0)\right] dk\\
	        & +\frac{1}{2\pi}\int _{0}^{2\pi}\left[A_{\phi}(k=-\pi)-A_{\phi}(k=\pi)\right] d\phi \\
	        &=\frac{1}{2\pi}\int_{-\pi}^{\pi}  dk\int_0^{2\pi} d\phi F_{k\phi}=C_1.\\
	      	\end{split}
\end{equation}
Above $A_{\phi}=\langle u_k(\phi)|i\partial_{\phi}|u_k(\phi)\rangle$ with $A_{\phi}(k=-\pi)=A_{\phi}(k=\pi)$ at the boundary of the first Brillouin Zone.

\begin{figure}[http]\centering
	\includegraphics[width=8.4cm]{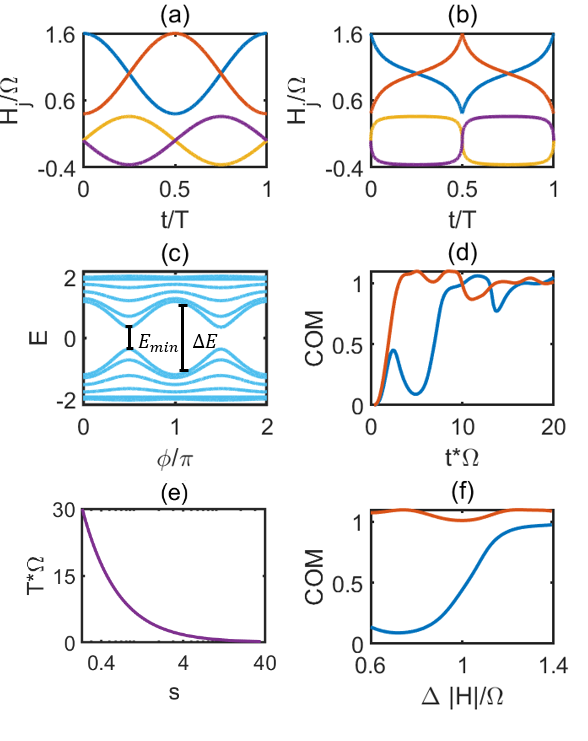}
	\caption{ Comparison of the linear and optimal modulations, and the system size $N=10$. The $t_j$($t_1$: blue line, $t_2$: red line, $t_A$: yellow line, $t_B$: purple line) vs $t/T$ for (a) the linear modulation and (b) the USE . (c) Energy spectrum with $\phi$ of this model.  The (d-e) show the transfer fidelity and the adjacent adiabatic coefficient $s$ as a function of $T*\Omega$ for the conventional linear modulation(blue line) and the optimal adiabatic modulation(red line) in the case of $\delta_0=0.6t_0$ and $\Delta_0=0.36t_0$. (f) The transfer fidelity vs hopping term variations for the conventional linear modulation(blue line) and the optimal adiabatic modulation(red line) in the case of $T*\Omega=6.8$.}
	\label{fig3}
\end{figure}

In realization experiments, the adiabatic parameter is linearly modulated as $\phi=\omega t$ and thus $\delta$ and $\Delta$ cyclically oscillate and vary with $t$ as seen in Fig \ref{fig3}(a). Conventional shortcut methods, including shortcuts to adiabaticity, quasiadiabatic dynamics and so on, cannot work in this model. We now also apply our shortcut theory to find the optimal adiabatic path of this pumping process. Due to the half-filled Rice-Mele model, there are only non-adiabatic transitions between the upper branch and lower branch of energy spectrum. Based on $E_{min}$ and $\Delta E$ that we chose as plotted in Fig.\ref{fig3}(c), the relevant parameter modulations are now modified by SUSM, as shown in Fig.\ref{fig3}(a,b). The fidelity vs evolution time with the linear modulation and SUSM are shown in Fig.\ref{fig3}(d) and the adjacent adiabatic coefficient $s$ is shown in Fig.\ref{fig3}(e) in the same condition. Moreover, Fig.\ref{fig3}(f) that the fidelity as a function of $t_j/\Omega$ for the linear modulation and SUSM. The result of this case, which is similar to the case of the 1D NH SSH model, shows that our scheme can reduce the evolution time and enhance the robustness without increasing the experimental complexity.

%

\emph{\color{blue}Conclusion and outlook.---}n summary, we propose a computationally efficient and experimentally friendly shortcut method for exact state manipulations in quantum topological systems. Theoretically, going beyond existing protocols for few-level systems, our shortcut only depends on a few pieces of information about the energy spectrum. Thanks to a stricter adiabatic condition, the spectrum-based shortcut protocol of the control parameter has been obtained. Experimentally, our control protocol does not need additional coupling or interaction, so this method is friendly. For example, our protocol shows great advantages in the standard Landau-Zener model, NH SSH model and Thouless pump model compared with the linear schemes or other shortcut medthods, e.g. faster rate of evolution and more robust against parameter error. Extending our ideas to the manipulations of corner states in higher-order topological insulators is intuitive. On the other hand, by incorporating nonlinear Landau-Zener  tunneling theory\cite{Wu,Wang,Cao}, we can readily extend our scheme to many-body systems, such as the Ising spin model.

	\begin{acknowledgements}
		This work was supported by the Natural Science Foundation of Guangdong Province under Grant 2023A1515011212 , the Special Projects in Key Fields of Ordinary Universities in Guangdong Province under Grant 2021ZDZX1015.
	\end{acknowledgements}

\end{document}